\documentclass[]{techport}

\usepackage{helvet}

\usepackage{amsmath}
\usepackage{natbib}
\usepackage{graphicx}
\usepackage{subcaption}

\usepackage[toc,page,header]{appendix}
\usepackage[utf8]{inputenc}
\usepackage[T1]{fontenc}
\usepackage{hyperref}
\usepackage{url}
\usepackage{booktabs}

\usepackage{amsfonts}
\usepackage{nicefrac}
\usepackage{microtype}
\usepackage{wrapfig}
\usepackage{amssymb}

\usepackage{titletoc}
\usepackage{minitoc}

\usepackage{geometry}
\usepackage{array}
\usepackage{etoolbox}

\definecolor{lightblue}{RGB}{200, 230, 255}
\definecolor{headerblue}{RGB}{150, 200, 255}

\usepackage{pgfplots}
\usepackage{xcolor}
\usepackage{float}
\usepackage{comment}
\usepackage{multirow}
\usepackage{multicol}
\usepackage{makecell}
\usepackage{arydshln}
\usepackage{siunitx}
\usepackage{tikz}
\usepackage{pgf-pie}
\usepackage[export]{adjustbox}

\usepackage{ragged2e}
\usepackage{xltabular}
\usepackage{tabularx}
\usepackage{caption}
\usepackage{enumitem}
\usepackage{pifont}
\usepackage[hang,flushmargin]{footmisc}

\usepackage{tcolorbox}
\tcbuselibrary{breakable}
\tcbuselibrary{skins}
\tcbuselibrary{theorems}
\tcbset{
  aibox/.style={
    width=\linewidth,
    top=8pt,
    bottom=4pt,
    colback=blue!6!white,
    colframe=black,
    colbacktitle=black,
    enhanced,
    center,
    attach boxed title to top left={yshift=-0.1in,xshift=0.15in},
    boxed title style={boxrule=0pt,colframe=white,},
  }
}
\newtcolorbox{AIbox}[2][]{aibox,title=#2,#1}

\usepackage{listings}
\usepackage{upquote}
\lstdefinelanguage{json}{
    basicstyle=\ttfamily\small,
    numbers=left,
    numberstyle=\tiny,
    stepnumber=1,
    numbersep=8pt,
    showstringspaces=false,
    breaklines=true,
    frame=single,
    backgroundcolor=\color{gray!5},
    string=[s]{"}{"},
    morestring=[b]',
    literate=
     *{:}{{{\color{blue}:}}}{1}
      {,}{{{\color{black},}}}{1}
      {"}{{{\color{red}\textquotedbl}}}{1},
}

\usepackage{threeparttable}
\usepackage{setspace}

\usepackage[scheme=plain]{ctex}
\usepackage[table]{xcolor}

\definecolor{oursgray}{gray}{0.95}

\definecolor{MossCyan}{HTML}{82D9FF} 
\definecolor{MossBlue}{HTML}{82B1FF}

\definecolor{tickG}{HTML}{00C853}
\definecolor{crossR}{HTML}{FF1744}

\newcommand{\cmark}{\textcolor{tickG}{\bfseries\ding{52}}}
\newcommand{\xmark}{\textcolor{crossR}{\bfseries\ding{56}}}
\newcommand{\PART}{\textcolor{orange}{\ding{52}\rotatebox[origin=c]{-9.2}{\kern-0.7em\ding{55}}}}

\newcommand{\hflogo}{\raisebox{-0.2ex}{\includegraphics[height=2.0ex]{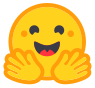}}}
\newcommand{\faGithub}{\raisebox{-0.2ex}{\includegraphics[height=2.0ex]{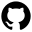}}}

\newtcolorbox{promptbox}[2][]{
    colback=white,
    coltext=black,
    arc=3mm,
    boxrule=0.5pt,
    colframe=black!60!white,
    title={#2},
    colbacktitle=black,
    coltitle=white,
    fonttitle=\bfseries,
    top=8pt,
    bottom=8pt,
    left=10pt,
    right=10pt,
    breakable,
    before upper={%
        \linespread{1}\selectfont
        \setlength{\parskip}{1ex plus 0.2ex minus 0.2ex}%
        \setlength{\parindent}{0pt}%
    },
    #1
}
\title{MMAE: A Massive Multitask Audio Editing Benchmark}

\author{
$^{1,2,3,4}$Ziyang Ma$^*$, $^{1,4}$Ruiqi Yan$^*$, $^{1}$Ruiyang Xu$^*$, $^{1}$Jie Fang$^*$, $^{1,2,4}$Zhikang Niu$^*$, $^{3}$Yi-Wen Chao$^*$, $^{1,4}$Wenming Tu$^*$, $^{3,4,5}$Tianrui Wang$^*$,
$^{4}$Auden, $^{1,2}$Qi Chen, $^{1,2}$Wenxi Chen, $^{1}$Jiaying Chi, $^{1}$Yanru Huo, $^{2}$Zixuan Jiang, $^{1}$Xiquan Li, $^{1}$Yalin Li, $^{1}$Junxi Liu, $^{6}$Minghao Liu, $^{1}$Binghao Qiang, $^{1}$Yijia Shan, $^{1}$Zheshu Song, $^{1}$Tian Tan, $^{1}$Zixiang Wang, $^{4,7}$Zeyu Xie, $^{3}$Zhifei Xie, $^{1}$Xiaoyu Xing, $^{1}$Qixiang Xu, $^{2,8}$Chen Yang, $^{1,2,4}$Guanrou Yang, $^{4}$Shan Yang, $^{1}$Yifan Yang, $^{4}$Steve Yves, $^{1}$Haotian Zhang, $^{1,2}$Haina Zhu, \\ 
$^{1}$Kai Yu, $^{4}$Liefeng Bo, $^{3}$Eng-Siong Chng, $^{1,2}$Xie Chen$^\dagger$ \\
    {\normalsize \normalfont
    $^{1}$Shanghai Jiao Tong University, $^{2}$Shanghai Innovation Institute, $^{3}$Nanyang Technological University \\
    $^{4}$Hunyuan Team, Tencent, $^{5}$Tianjin University, $^{6}$ZODA, $^{7}$Peking University, $^{8}$Fudan University \\
    }
}

\logo[width=3.5cm]{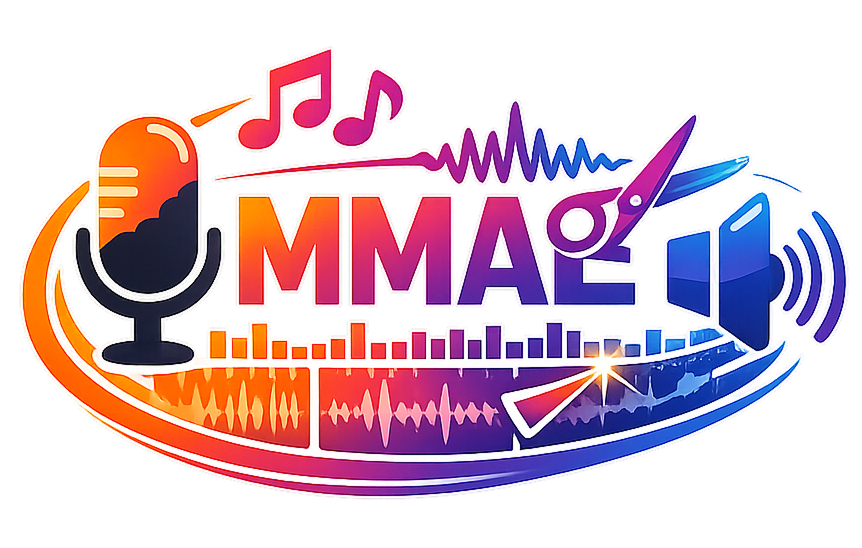}
\renewcommand{\logoskipbefore}{0mm}  
\renewcommand{\logoskipafter}{0mm} 

\checkdata[\raisebox{-0.1ex}{\faGithub}\hspace{0.4em}GitHub]{\url{https://github.com/ddlBoJack/MMAE}}
\checkdata[\raisebox{-0.1ex}{\hflogo}\hspace{0.4em}Hugging Face]{\url{https://huggingface.co/datasets/BoJack/MMAE}}

{\let\thefootnote\relax\footnotetext{$^*$Core Contributors. Other authors are listed in surname alphabetical order.}}
{\let\thefootnote\relax\footnotetext{$^\dagger$Corresponding Author.}}

\abstract{
We introduce \textbf{MMAE}, a \textbf{M}assive \textbf{M}ultitask \textbf{A}udio \textbf{E}diting benchmark, serving as the first comprehensive evaluation testbed designed for general-purpose instruction-based audio editing. 
Spurred by the shift toward intelligent creation, interactive editing has rapidly expanded from visual domains, pioneered by models like Nano-banana 2 for images and Gemini-Omni for video, into audio. 
However, the current evaluation infrastructure lags severely, remaining highly fragmented and restricted to specific subdomains or basic operations. Unlike existing benchmarks that are limited in scope, MMAE extends to a broad spectrum of real-world scenarios, encompassing 7 distinct audio modalities, including sound, speech, music, and their mixtures. Furthermore, we establish a comprehensive taxonomy spanning 6 levels of task complexity, from basic modifications to multi-hop reasoning and multi-round editing, 2 levels of granularity, and 8 distinct operation types. Meticulously curated through human-agent collaboration, MMAE comprises 2,000 high-fidelity samples paired with a pioneering rubric-based evaluation framework. By decomposing free-form tasks into 17,741 verifiable criteria, this robust rubric-based paradigm enables a precise, multi-dimensional assessment of both instruction following and context consistency. Our extensive evaluation of leading models reveals that current systems remain far from achieving reliable edits. Strikingly, the Exact Match Rate (EMR) consistently falls below 5\% and plummets to an absolute 0\% in complex, mixed-modality tasks, exposing critical bottlenecks in precise execution and structural robustness. We hope MMAE will serve as a catalyst for future advances in the intelligent creation community, providing a clear diagnostic roadmap and establishing a standardized, long-lasting evaluation paradigm for next-generation audio editing systems. 
}

\begin{document}
\maketitle

\begingroup
\renewcommand{\thefootnote}{\fnsymbol{footnote}}
\endgroup

\newpage

\begin{figure}[H]
    \centering
    \includegraphics[width=\textwidth]{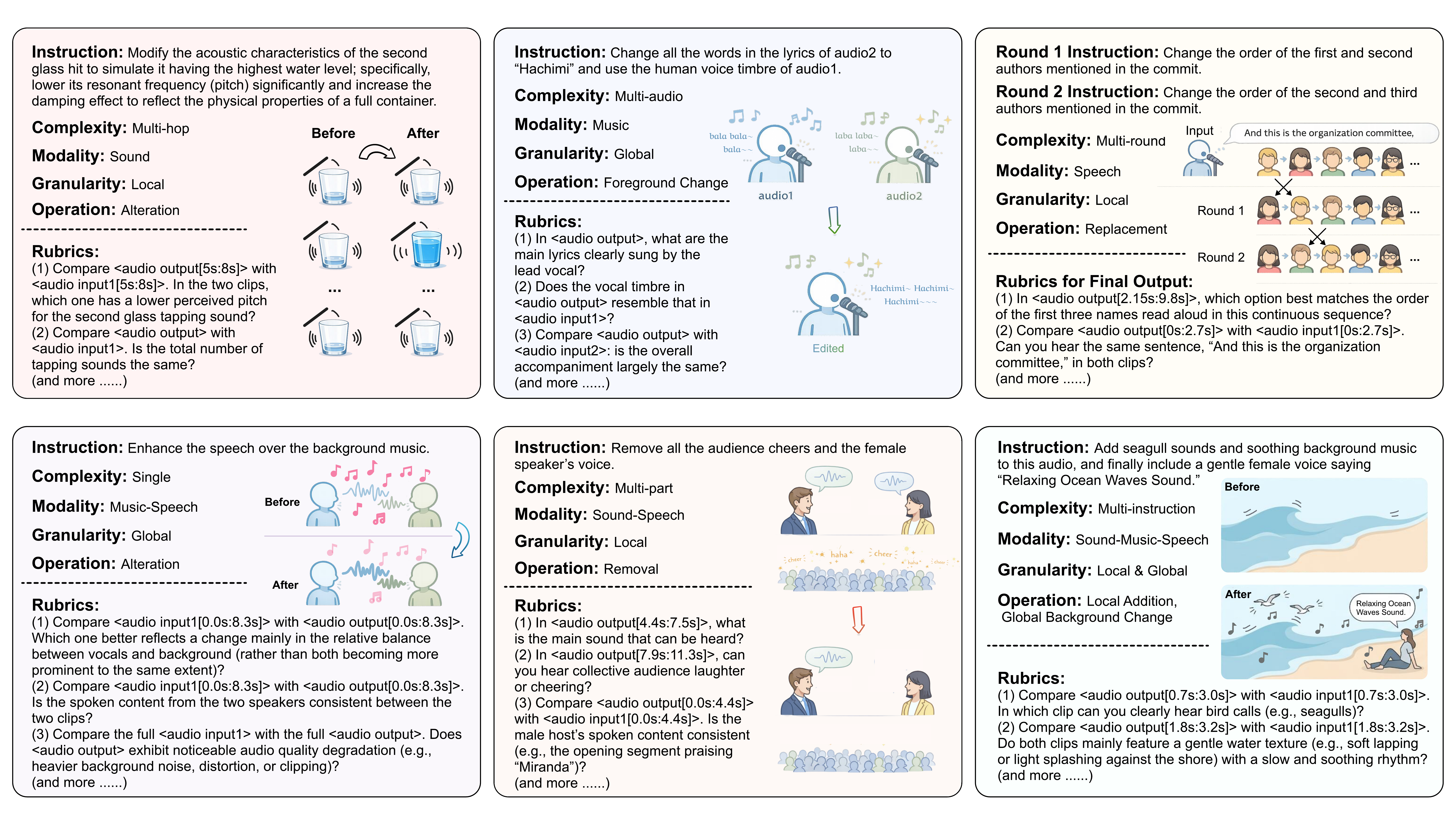}
    \caption{Examples from the MMAE benchmark, illustrating the overall taxonomy and the proposed rubric-based evaluation paradigm. These examples span diverse modalities, complexity, and operation types, with a subset of associated rubrics shown for clarity.}
    \label{fig:examples}
\end{figure}

\section{Introduction}
Intelligent editing has witnessed remarkable breakthroughs in recent years, transitioning generative models from single-element editing to interactive manipulation. 
In the visual domain, image editing applications like Nano-banana 2~\cite{google_nano_banana2} have successfully matured into practical production workflows, while recent advancements in video editing models like Gemini-Omni~\cite{google_gemini_omni} have vastly expanded the creative boundaries. 
Spurred by this shift toward interactive intelligence, the audio community has recently seen a surge of instruction-based audio editing models~\cite{lan2025SmartDJ, tao2025mmedit, yan2025Ming-UniAudio, yan2025Step-Audio-EditX, chen2026audiochat, tian2026Audio-Omni}. By allowing users to alter speech, music, or sound effects via open-ended natural language instructions, these models represent the next-generation paradigm for intelligent audio generation and editing systems. 

However, the current evaluation infrastructure for audio editing lags severely behind. 
To effectively assess these rapidly evolving systems, a next-generation evaluation framework must achieve breakthroughs in two critical dimensions: 1) data coverage and 2) evaluation paradigm. 
First, existing benchmarks~\cite{peng2024voicecraft, yan2025Ming-UniAudio, yan2025Step-Audio-EditX, chen2026audiochat} are highly fragmented, typically restricted to specific subdomains (e.g., speech-only or sound-only) or basic operations (e.g., addition, removal, replacement). A modern testbed must comprehensively cover speech, music, and sound effects, while featuring complex scenarios designed to stress-test a model's entire cognitive pipeline from nuanced perception and implicit reasoning to high-fidelity generation. 
Second, traditional metrics have proven inadequate for open-ended instructional editing. To establish a robust and reliable evaluation paradigm, rubric-based frameworks have recently been validated in text reinforcement learning reward~\cite{gunjal2025rubrics-as-rewards}, audio reasoning~\cite{ma2026interspeech-mmar}, and image editing~\cite{bai2026Edit-Compass}, which successfully decompose free-form, multifaceted tasks into structured and verifiable criteria. Extending this proven paradigm presents a highly effective solution for evaluating complex, instruction-based audio editing. 

To bridge this critical gap, we introduce \textbf{MMAE}, a \textbf{M}assive \textbf{M}ultitask \textbf{A}udio \textbf{E}diting benchmark, serving as the first comprehensive evaluation benchmark designed for general-purpose instruction-based audio editing. 
As illustrated in Figure~\ref{fig:examples}, MMAE encompasses challenging, creative, and practical editing tasks, pairing each audio-instruction instance with a tailored set of rubrics to precisely quantify the editing outcome. Specifically, we establish a comprehensive taxonomy spanning 7 types of audio modalities (e.g., sound, speech, music, and their mixtures), 6 levels of task complexity ranging from simple modifications to complex demands (e.g., multi-hop reasoning and context-aware multi-round editing), 2 levels of granularity covering both local and global edits, and 8 distinct operation categories. 
Besides, the novel rubric-based paradigm enables a fine-grained, multi-dimensional assessment of editing correctness and generation quality, maximizing both reliability and interpretability to effectively diagnose model capabilities. 
Finally, the data curation pipeline of MMAE is rigorously designed, integrating heuristic audio collection, human-agent collaborative annotation, multi-stage refinement, and strict manual quality control. This extensive effort yields a high-fidelity benchmark comprising 2,000 diverse samples and 17,741 meticulously crafted rubrics. 

Based on the MMAE benchmark, our evaluation of the 5 latest audio editing models reveals that while current systems possess basic capabilities, they remain far from achieving reliable and flawless edits. Strikingly, the Exact Match Rate (EMR) consistently falls below 5\% across all models, even plummeting to an absolute 0\% in complex, mixed-modality scenarios. 
Our detailed analysis highlights several critical limitations driving this failure: models fundamentally struggle to balance precise instruction execution with the strict preservation of unrelated acoustic contexts; they exhibit a severe lack of structural robustness as task complexity and cross-domain synchronization demands increase; and they reveal a clear decoupling between average metric competency and flawless execution. Furthermore, incorporating external agentic planners yields no consistent improvement, and we found critical bottlenecks on both the understanding and generation sides. 

In summary, our main contributions are:
\begin{itemize}
    \item We introduce MMAE, the first universal benchmark for evaluating instruction-based audio editing capabilities, aiming to establish a standardized evaluation paradigm for next-generation audio editing systems. 
    \item We construct a comprehensive, scalable evaluation suite with a systematic taxonomy, high-quality annotations, and well-defined rubric-based metrics, enabling rigorous assessment of audio editing performance.
    \item We benchmark a wide range of leading audio editing models, exposing critical bottlenecks in current models or systems. Our empirical insights provide a clear diagnostic roadmap to guide the development of more advanced and robust next-generation systems.
\end{itemize}
\section{Related Work}

\subsection{Audio Editing Models}

Audio editing has been studied in relatively constrained settings in earlier works~\cite{jiang2023fluentspeech, wang2025Ssr-speech, ellis2025recomposer, xu2024Prompt-guided-precise-audio-editing, manor2024Zero-shot-unsupervised-and-text-based-audio-editing, huang2024instructspeech, liang2024wavcraft, wang2023audit, jia2025audioeditor, liang2025audiomorphix, peng2024voicecraft, zheng2025VoiceCraft-X}, typically limited to specific modalities or predefined operation types. More recently, a growing body of research~\cite{lan2025SmartDJ, tao2025mmedit, yan2025Ming-UniAudio, yan2025Step-Audio-EditX, chen2026audiochat, tian2026Audio-Omni} has emerged along several dimensions, such as natural language instruction-guided editing, cross-modal and multi-domain generalization, and open-ended compositional editing, reflecting a broader trend toward versatile and unified audio editing systems.

Early exploration into general audio (specifically, sound effects) editing was pioneered by AUDIT~\cite{wang2023audit}, which introduced a latent diffusion model trained on synthetic triplets to perform text-guided addition, removal, and replacement of sound events. AudioEditor and AudioMorphix~\cite{jia2025audioeditor, liang2025audiomorphix} subsequently demonstrated that comparable editing capabilities could be achieved in a training-free method. Moving beyond basic event manipulation, MMEdit~\cite{tao2025mmedit} expanded the task scope to more diverse operations by leveraging an audio language model for joint source-instruction understanding. Alternatively, SmartDJ~\cite{lan2025SmartDJ} introduced a hierarchical two-stage framework, where an audio language model decomposes high-level declarative instructions into atomic steps, which are then executed sequentially by a diffusion-based editor.
In the speech domain, VoiceCraft~\cite{peng2024voicecraft} proposed a neural codec language model with a novel token rearrangement procedure for zero-shot speech content editing. CosyEdit~\cite{chen2026cosyedit} demonstrated that speech editing can be unlocked from pretrained zero-shot TTS models via lightweight post-training with only a little amount of data. Recent advances further push the boundaries of expressiveness and integration: Step-Audio-EditX~\cite{yan2025Step-Audio-EditX} introduces nuanced control over emotion and paralinguistic attributes, while Ming-UniAudio~\cite{yan2025Ming-UniAudio} unifies speech understanding, generation, and free-form editing within a single framework via a continuous VAE tokenizer. 

Several recent systems pursue broader unification across diverse audio domains. Audio-Omni~\cite{tian2026Audio-Omni} combines a frozen multimodal LLM with a diffusion transformer to handle generation and editing across sound, music, and speech simultaneously. AudioChat~\cite{chen2026audiochat} achieves unified understanding, generation, and editing through a novel transfusion forcing objective. Furthermore， InstructAV2AV~\cite{zheng2026instructav2av} and SpongeBob~\cite{liang2026spongebob} extend the scope to joint audio-visual editing with synchronized cross-modal generation.

\subsection{Audio Editing Evaluations}
Despite the rapid emergence of audio editing models, a comprehensive and dedicated evaluation benchmark for this field remains absent. Current evaluation efforts are highly fragmented and strictly domain-specific. In the speech domain, existing test sets like RealEdit~\cite{peng2024voicecraft} focus on localized manipulation (insertion, deletion, and substitution) evaluated via word error rate (WER) and speaker similarity. While newer benchmarks such as Ming-Freeform-Audio-Edit~\cite{yan2025Ming-UniAudio} and Step-Audio-Edit-Benchmark~\cite{yan2025Step-Audio-EditX} expand into semantic, acoustic, and expressive attributes, they remain confined to a restricted set of task archetypes that lacks broader operational diversity. For general audio, evaluation provided by MMEdit~\cite{tao2025mmedit} and Audio-Omni~\cite{tian2026Audio-Omni} is also limited to a narrow set of basic operations, whereas StoryGen-Eval~\cite{chen2026audiochat} targets multi-source storytelling using FLAM-based~\cite{wu2025flam} metrics. 
Crucially, these evaluation sets are each restricted to a narrow subset of editing tasks within a single domain, and typically rely on traditional signal-level metrics (e.g., FAD, LSD, CLAP similarity) or generic MOS ratings, failing to explicitly assess editing correctness.

To bridge this critical gap, MMAE serves as the first comprehensive benchmark designed for universal audio editing evaluation across all audio modalities, including sound, music, speech, and their mix. MMAE establishes a systematic taxonomy spanning modality, complexity, and operation dimensions. Moving beyond synthetic data, it features diverse, real-world audio samples paired with human-annotated natural language instructions. Furthermore, instead of relying on coarse, insufficiently robust metrics, MMAE employs fine-grained, rubric-based evaluation where each sample is assessed through multiple atomic, objective criteria targeting both instruction following and content consistency. This design enables precise, interpretable, and universal measurement of editing quality, providing a robust diagnostic tool for next-generation audio editing models.
\section{MMAE}

\subsection{Overview}
MMAE is designed to evaluate next-generation, instruction-based intelligent audio editing systems. Moving beyond simple execution, it demands that models seamlessly integrate three core capabilities: \textbf{perception} (to understand the source audio context), \textbf{reasoning} (to interpret complex, implicit user intent), and \textbf{generation} (to execute high-fidelity edits). 

MMAE comprises 2k samples paired with over 17k fine-grained rubrics, covering challenging, creative, and practical editing tasks that require models to handle complex transformations across a wide range of editing scenarios. Each sample features an open-ended natural language instruction spanning diverse audio modalities, task complexity, and editing operations. The benchmark is constructed through a rigorous pipeline, where initial annotations are generated via human-agent collaboration and subsequently refined, validated, and quality-controlled through expert review. This workflow ensures both diversity and high quality, making MMAE a reliable benchmark for advancing the field.

\subsection{Taxonomy}

\begin{figure*}[t!]
    \centering
    \begin{subcaptiongroup}
    \begin{subfigure}[t]{0.33\linewidth}
        \centering
        \includegraphics[width=\linewidth]{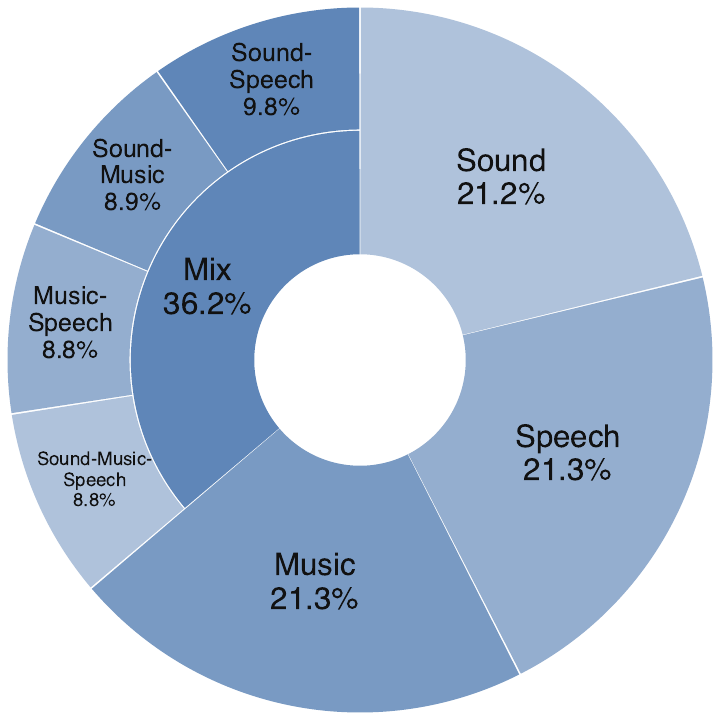}
        \caption{Modality distribution}
        \label{fig:pie_modality}
    \end{subfigure}
    \hfill
    \begin{subfigure}[t]{0.33\linewidth}
        \centering
        \includegraphics[width=\linewidth]{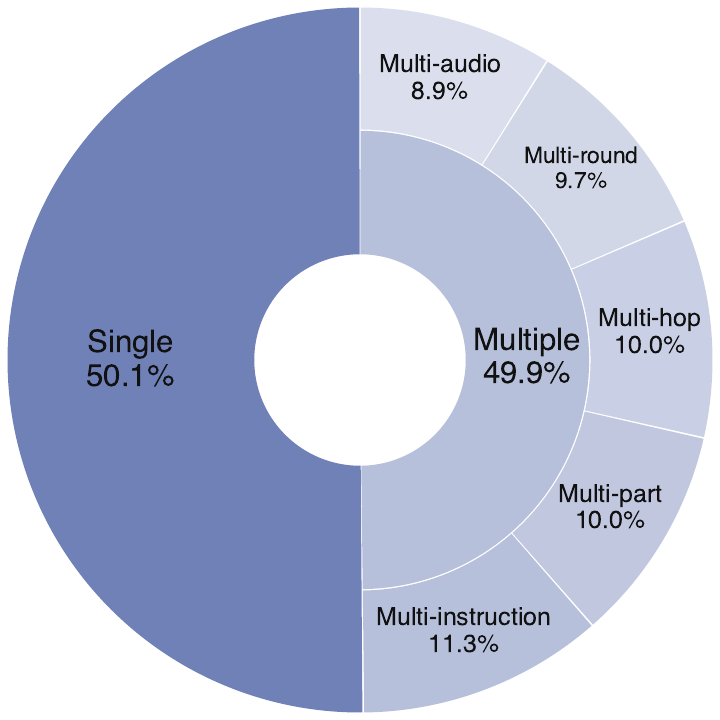}
        \caption{Complexity distribution}
        \label{fig:pie_difficulty}
    \end{subfigure}
    \hfill
    \begin{subfigure}[t]{0.33\linewidth}
        \centering
        \includegraphics[width=\linewidth]{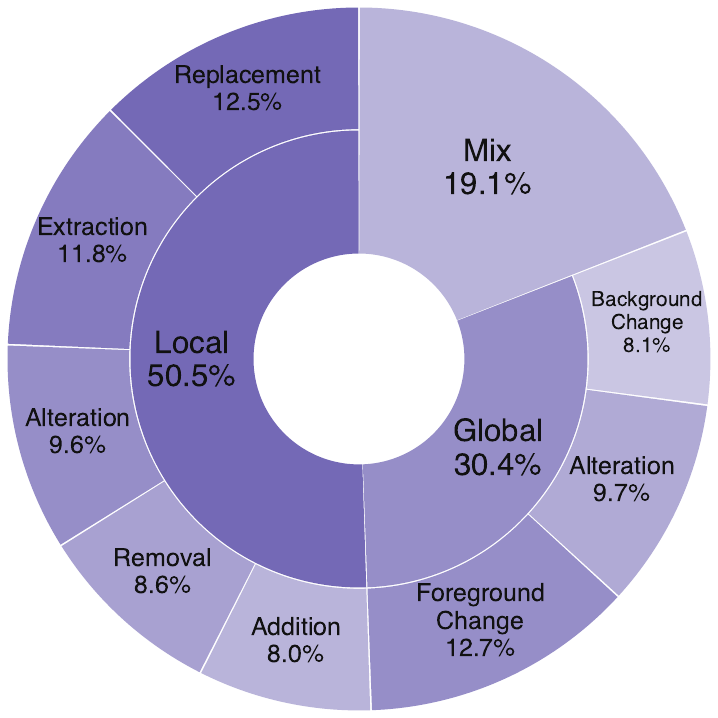}
        \caption{Operation distribution}
        \label{fig:pie_operation}
    \end{subfigure}
    \end{subcaptiongroup}
    \caption{Distribution of the MMAE benchmark across three taxonomy dimensions.}
    \label{fig:pie_distribution}
\end{figure*}

We design a parallel taxonomy to systematically characterize audio editing tasks from three orthogonal dimensions: \textbf{Modality}, \textbf{Complexity}, and \textbf{Operation}. 
These dimensions are defined in a compositional manner, enabling flexible combinations to cover a wide spectrum of multitask audio editing scenarios. Figure~\ref{fig:pie_distribution} presents the overall distribution of MMAE.

\paragraph{\textbf{Modality.}}
The modality dimension captures the types of audio involved in the editing task. 
Considering the diverse and often mixed-modality nature of real-world audio, we focus on seven categories: \textit{sound}, \textit{music}, \textit{speech}, and their combinations, including \textit{sound-music}, \textit{sound-speech}, \textit{music-speech}, and \textit{sound-music-speech}. 

\paragraph{\textbf{Complexity.}}
The complexity dimension characterizes the structural and cognitive difficulty of the task, which clearly reflects the cases of realistic usage scenarios. Tasks are stratified into six levels:
\begin{itemize}
    \item \textit{Single}: basic editing tasks involving a single operation on a single element;
    \item \textit{Multi-part}: tasks with a single instruction that involves multiple elements;
    \item \textit{Multi-instruction}: samples with commands composed of multiple independent single instructions;
    \item \textit{Multi-audio}: tasks involving multiple audio sources to accomplish; 
    \item \textit{Multi-round}: iterative editing across multiple turns, where later edits depend on earlier ones;
    \item \textit{Multi-hop}: tasks requiring multi-hop reasoning or intermediate inference to determine the expected output. 
\end{itemize}

\paragraph{\textbf{Operation.}}
The operation dimension describes the types of editing actions, organized by granularity into \textit{local} and \textit{global} edits. 
\textit{Local} edits focus on specific segments or elements within the audio, including \textit{addition}, \textit{removal}, \textit{replacement}, \textit{extraction}, and \textit{alteration} (e.g., modifying localized attributes such as timbre or pitch). 
\textit{Global} edits operate on the entire audio or its overall characteristics, including \textit{background change}, \textit{foreground change}, and \textit{global alteration}.
Depending on the instruction, each sample may involve either a single operation or an arbitrary composition of multiple operations.

\subsection{Evaluation Paradigm}
Evaluating instruction-based audio editing requires a paradigm capable of benchmarking both editing correctness and generation quality. Existing metrics fall short when confronted with complex, open-ended multi-modal editing. To address this, we introduce a structured evaluation framework anchored in an \textbf{instance-level, rubric-based paradigm}. By decomposing multifaceted editing tasks into localized, verifiable checkpoints, this approach provides an objective and interpretable mechanism to diagnose model performance. 

\paragraph{\textbf{Evaluation Dimensions.}}
To rigorously assess the multi-faceted nature of audio editing, we evaluate models along two core, complementary dimensions:
\begin{itemize}
    \item \textit{Instruction Following} measures editing execution accuracy, specifically, whether the model precisely performs the modifications requested by the natural language instruction.
    \item \textit{Consistency} evaluates context preservation, ensuring that all acoustic elements irrelevant to the editing command remain strictly unaltered. 
\end{itemize}
Together, these dimensions capture the fundamental trade-off in audio editing: executing precise modifications or global transformations while maintaining the holistic fidelity of the original audio. 

\paragraph{\textbf{Rubric-Based Metrics.}}
To enable fine-grained and objective evaluation, we adopt a rubric-based evaluation paradigm. 
For each audio editing sample, we design a set of comprehensive, atomic, and mutually independent evaluation criteria (rubrics), which assess model performance from multiple perspectives. 
Each rubric is formulated as a multiple-choice question, where one correct option corresponds to successful editing behavior, and the other incorrect options indicate failure. 
An external judger, instantiated as a high-performance audio language model, is responsible for selecting the appropriate option based on the editing output. And each rubric is assigned a binary score (1 or 0) based on whether the option selected by the judger matches the correct option.
Our rubric design is governed by four foundational principles:
\begin{itemize}
    \item \textit{Completeness}: The evaluation criteria should cover all relevant aspects of the editing task to avoid missing important factors. 
    \item \textit{Atomicity}: Each rubric should focus on a single, indivisible property (e.g., duration, timbre, background sound, spoken content, etc.) that falls well within the perceptual and evaluative capabilities of the judge model. 
    \item \textit{Orthogonality}: Different rubrics should be independent, such that the outcome of one criterion does not directly imply another. 
    \item \textit{Objectivity}: Rubrics should be defined based on observable and verifiable properties, minimizing ambiguity and subjective judgment, thereby enabling the judger to function as a reliable measurement instrument.
\end{itemize}

\begin{table*}[t!]
    \centering
    \caption{Key statistics of the MMAE benchmark.}\label{tab:statistics}
    \begin{minipage}[t]{0.48\linewidth}
        \centering
        \resizebox{\linewidth}{!}{
        \begin{tabular}{lc}
            \toprule
            \textbf{Statistics} & \textbf{Number} \\
            \midrule
            Total Samples & 2,000 \\
            Total Rubrics & 17,741 \\
            Avg. Operations / Sample & 1.22 \\
            Avg. Audio Duration / Sample & 14.46 sec \\
            Avg. Instruction Length & 14.00 words \\
            \bottomrule
        \end{tabular}
        }
    \end{minipage}
    \hfill
    \begin{minipage}[t]{0.48\linewidth}
        \centering
        \resizebox{\linewidth}{!}{
        \begin{tabular}{lc}
            \toprule
            \textbf{Statistics} & \textbf{Number} \\
            \midrule
            Avg. Rubrics / Sample & 8.87 \\
            Avg. IF Rubrics / Sample & 3.58 \\
            Avg. Consistency Rubrics / Sample & 5.29 \\
            Avg. Choices / Rubric & 3.53 \\
            Avg. Rubric Question Length & 25.45 words \\
            \bottomrule
        \end{tabular}
        }
    \end{minipage}
\end{table*}

\paragraph{\textbf{Statistics.}}
As detailed in Table~\ref{tab:statistics}, MMAE comprises 2,000 samples and 17,741 rubrics, averaging 8.87 rubrics per sample (3.58 for Instruction Following and 5.29 for Consistency). On average, each sample spans 14.46 seconds and contains 1.22 editing operations, guided by a 14-word instruction. Furthermore, the evaluation questions average 25.45 words with 3.53 choices per rubric, ensuring a highly discriminative setup for comprehensive assessment.

\subsection{Data Curation Pipeline}

\begin{figure*}[t!]
    \centering
    \includegraphics[width=\linewidth]{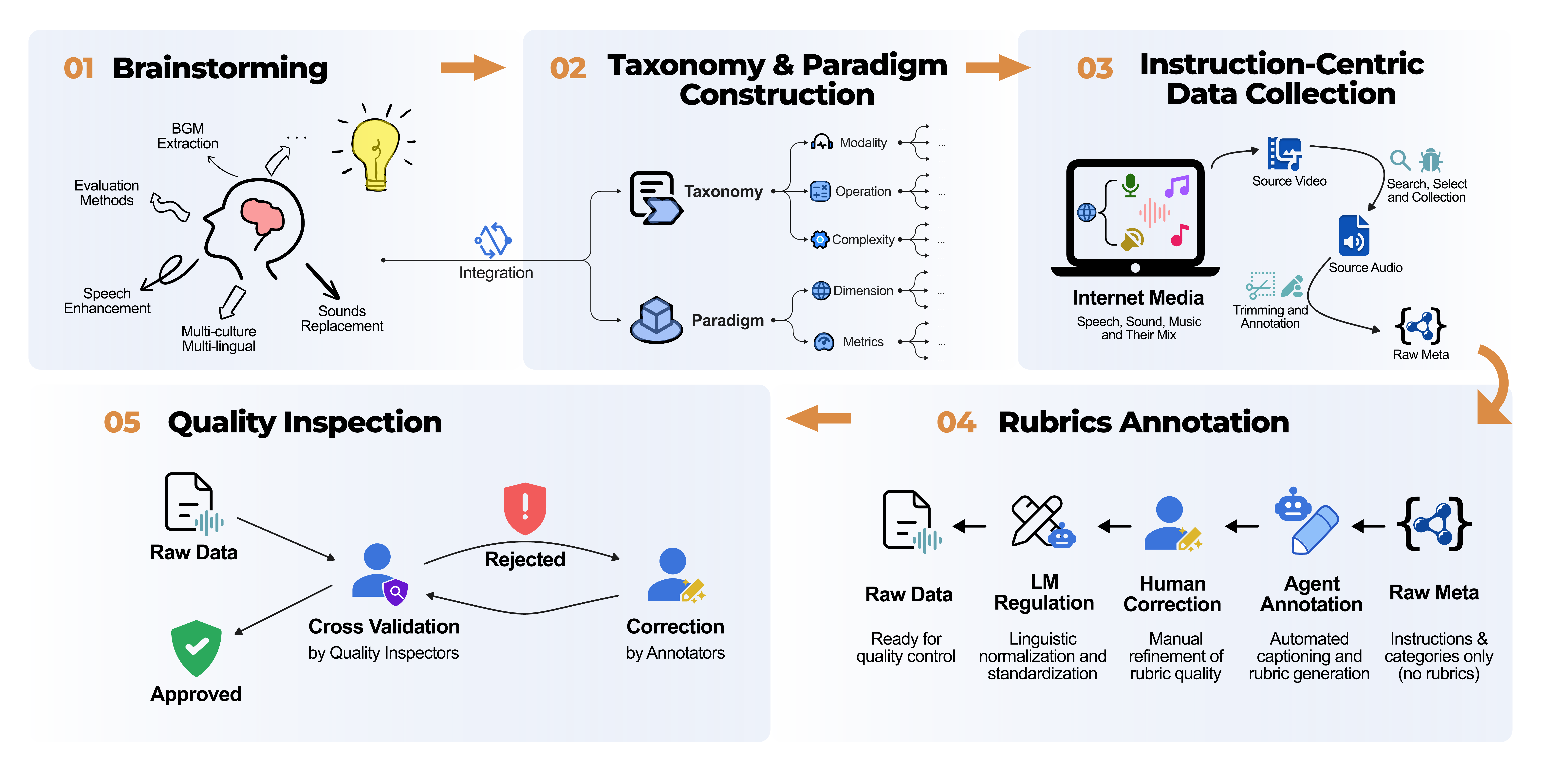}
    \caption{A comprehensive data curation pipeline of the MMAE benchmark. The process includes: (1) expert-driven brainstorming to collect diverse audio editing scenarios; (2) taxonomy and paradigm construction, establishing the multi-dimensional task taxonomy and the rubric-based evaluation framework; (3) instruction-centric data collection with dynamic balancing across taxonomy dimensions; (4) human-agent collaborative annotation with automated rubric generation and human refinement; and (5) iterative quality inspection with revision and filtering to ensure data quality.}
    \label{fig:pipeline}
\end{figure*}

As illustrated in Figure~\ref{fig:pipeline}, MMAE is constructed through a systematic five-stage pipeline designed to ensure both diversity and high-quality of the benchmark:

\textbf{1) Brainstorming.} 
We organize multiple rounds of brainstorming sessions with expert annotators to collect diverse audio editing ideas. 
Given the open-ended nature of audio editing, this stage focuses on gathering a wide range of intuitive, creative, and practical editing scenarios, covering different modalities and levels of complexity.

\textbf{2) Taxonomy \& Paradigm Construction.} 
Based on the brainstorming outcomes, we consolidate the core design of the benchmark, including the task taxonomy and the evaluation paradigm. 
In particular, we establish the rubric-based evaluation framework and the orthogonal taxonomy spanning modality, complexity, and operation. 
This stage provides a systematic foundation for subsequent data construction. 

\textbf{3) Instruction-Centric Data Collection.} 
Annotators manually search and collect audio data from online videos, including retrieving raw audio and trimming it into input clips. 
For each instance, annotators write instructions and label relevant metadata, including audio modality, task complexity, operation types, and keywords. 
To ensure balanced coverage across different task types, we adopt a dynamic balancing strategy along the three core taxonomy dimensions, resulting in a diverse and well-distributed dataset.

\textbf{4) Rubrics Annotation.} 
We employ a human-agent collaborative workflow to efficiently construct evaluation rubrics. 
To mitigate description hallucinations and maximize caption precision, we leverage the Omni-Detective~\cite{ma2025omni} agentic pipeline to extract detailed audio captions from the raw clips, which has been proven effective in the Qwen-Omni series~\cite{xu2025Qwen3-omni, team2026qwen3}. 
An LLM is then utilized to generate initial rubric drafts from detailed captions, user instructions, and metadata. 
Subsequently, human annotators refine these rubrics by adding, removing, or refining individual rubric items. 
Finally, an LLM is used for post-processing, including normalization and standardization of rubric expressions. 
This human-agent collaborative workflow significantly improves both annotation efficiency and effectiveness. 

\textbf{5) Quality Inspection.} 
Strict quality control is enforced through a dedicated cross-review protocol. 
Each data item is independently reviewed by blind inspectors who have no prior exposure. 
Samples that fail quality checks are iteratively revised until they pass the acceptance criteria, while irreparable cases are discarded to guarantee a high-fidelity final dataset. 
\begin{table*}[t!]
\centering
\caption{
Main results on the MMAE benchmark. (a) Performance grouped by complexity, reporting scores for \textit{single} and \textit{multiple} categories, along with the overall score. (b)(c) Performance breakdown across different modalities, with scores reported within each category. IFR = Instruction Following Rate, CR = Consistency Rate, EMR = Exact Match Rate. The best results are presented in \textbf{bold}. ``Identity" denotes directly returning the input without modification, while ``Noise" denotes generating pure noise. Results under these settings are reported as baselines for reference. $^*$MMEdit, Audio-Omni, and SmartDJ are either limited to inputs of at most 10 seconds or trained solely on data with durations $\leq10$ seconds. Accordingly, we evaluate these models only on samples with duration $\leq10$ seconds (801 samples).
}
\label{tab:main_results}

\subcaption{Results by complexity.}\label{tab:result_complexity}
\begin{tabular}{l ccc ccc ccc}
\toprule
\toprule
\multirow{2.5}{*}{\textbf{Model}} & \multicolumn{3}{c}{\textbf{Single}} & \multicolumn{3}{c}{\textbf{Multiple}} & \multicolumn{3}{c}{\textbf{Overall}} \\
\cmidrule(lr){2-4} \cmidrule(lr){5-7} \cmidrule(lr){8-10}
& IFR & CR & EMR & IFR & CR & EMR & IFR & CR & EMR \\
\midrule

Identity & 23.87 & 96.15 & 2.89 & 30.90 & 92.09 & 6.32 & 27.37 & 94.13 & 4.60 \\
Noise & 35.03 & 15.23 & 0.00 & 29.12 & 16.14 & 0.00 & 32.08 & 15.68 & 0.00 \\

\midrule \midrule

Step-Audio-EditX & \textbf{46.64} & \textbf{59.06} & 3.99 & \textbf{43.06} & \textbf{58.69} & \textbf{2.11} & \textbf{44.86} & \textbf{58.88} & 3.05 \\
Ming-UniAudio & 31.74 & 53.83 & \textbf{4.59} & 27.90 & {51.57} & 1.81 & 29.82 & 52.71 & \textbf{3.20} \\
\cdashline{1-10}
\noalign{\vskip 1.0mm}
MMEdit$^*$ & {48.39} & 54.27 & 4.86 & 36.94 & 39.86 & 1.90 & 43.12 & 47.64 & 3.50 \\
Audio-Omni$^*$ & \textbf{58.43} & \textbf{64.57} & \textbf{6.25} & \textbf{41.70} & \textbf{47.94} & \textbf{3.52} & \textbf{50.73} & \textbf{56.93} & \textbf{4.99} \\
SmartDJ$^*$ w/o planer & 42.52 & {63.91} & {5.56} & 33.12 & 45.43 & \textbf{3.52} & 38.20 & 55.41 & {4.62} \\
SmartDJ$^*$ w/ planer & 47.54 & 55.09 & 3.47 & 36.06 & 40.38 & {2.71} & 42.26 & 48.33 & 3.12 \\

\bottomrule
\bottomrule
\end{tabular}

\vspace{6mm}

\subcaption{Results on single modality.}\label{tab:result_single_modality}
\begin{tabular}{l ccc ccc ccc}
\toprule
\toprule
\multirow{2.5}{*}{\textbf{Model}} & \multicolumn{3}{c}{\textbf{Sound}} & \multicolumn{3}{c}{\textbf{Music}} & \multicolumn{3}{c}{\textbf{Speech}} \\
\cmidrule(lr){2-4} \cmidrule(lr){5-7} \cmidrule(lr){8-10}
& IFR & CR & EMR & IFR & CR & EMR & IFR & CR & EMR \\
\midrule

Identity & 30.40 & 94.44 & 6.37 & 25.59 & 94.42 & 4.93 & 31.24 & 92.59 & 3.99 \\
Noise & 40.78 & 20.25 & 0.00 & 31.45 & 13.89 & 0.00 & 20.90 & 14.72 & 0.00 \\

\midrule \midrule

Step-Audio-EditX & \textbf{46.51} & \textbf{51.38} & \textbf{3.07} & \textbf{42.75} & \textbf{47.84} & \textbf{1.88} & \textbf{43.52} & \textbf{77.27} & {4.69} \\
Ming-UniAudio & 28.77 & 47.88 & 2.12 & 28.97 & 34.71 & 0.94 & 34.13 & {76.01} & \textbf{7.04} \\
\cdashline{1-10}
\noalign{\vskip 1.0mm}
MMEdit$^*$ & {53.07} & 54.22 & 4.88 & 41.85 & 47.19 & 3.94 & 30.52 & 35.40 & 0.99 \\
Audio-Omni$^*$ & \textbf{56.58} & {56.02} & \textbf{7.72} & \textbf{56.97} & {52.55} & {5.51} & \textbf{43.14} & \textbf{68.29} & 1.98 \\
SmartDJ$^*$ w/o planer & 48.12 & \textbf{64.45} & {6.50} & 38.50 & \textbf{56.88} & \textbf{7.09} & 28.03 & 56.22 & \textbf{2.97} \\
SmartDJ$^*$ w/ planer & 51.74 & 53.19 & 4.88 & {46.17} & 46.06 & 2.36 & 32.17 & 43.00 & 0.99 \\

\bottomrule
\bottomrule
\end{tabular}

\vspace{6mm}

\subcaption{Results on mixed modality.}\label{tab:result_mix_modality}
\resizebox{\linewidth}{!}{
\begin{tabular}{l ccc ccc ccc ccc}
\toprule
\toprule
\multirow{2.5}{*}{\textbf{Model}} & \multicolumn{3}{c}{\textbf{Sound-Music}} & \multicolumn{3}{c}{\textbf{Sound-Speech}} & \multicolumn{3}{c}{\textbf{Music-Speech}} & \multicolumn{3}{c}{\textbf{Sound-Music-Speech}} \\
\cmidrule(lr){2-4} \cmidrule(lr){5-7} \cmidrule(lr){8-10} \cmidrule(lr){11-13}
& IFR & CR & EMR & IFR & CR & EMR & IFR & CR & EMR & IFR & CR & EMR \\
\midrule

Identity & 21.16 & 94.79 & 2.23 & 28.11 & 96.15 & 7.69 & 25.82 & 95.71 & 2.86 & 22.08 & 91.92 & 1.71 \\
Noise & 34.09 & 18.14 & 0.00 & 39.03 & 12.42 & 0.00 & 32.15 & 11.88 & 0.00 & 29.97 & 16.18 & 0.00 \\

\midrule \midrule

Step-Audio-EditX & \textbf{41.30} & \textbf{44.61} & 0.00 & \textbf{48.79} & \textbf{66.61} & 5.13 & \textbf{48.66} & \textbf{65.98} & \textbf{4.00} & \textbf{44.73} & \textbf{58.47} & 1.71 \\
Ming-UniAudio & 27.57 & 34.94 & 0.00 & 29.23 & {65.84} & \textbf{6.15} & 29.84 & 57.37 & 2.29 & 26.93 & {51.02} & \textbf{2.86} \\
\cdashline{1-13}
\noalign{\vskip 1.0mm}
MMEdit$^*$ & {49.29} & 48.04 & {4.92} & \textbf{46.15} & 53.25 & 5.08 & 44.04 & \textbf{60.24} & \textbf{3.85} & 36.84 & 46.25 & 1.85 \\
Audio-Omni$^*$ & \textbf{57.32} & 43.41 & 3.28 & 43.68 & \textbf{54.47} & \textbf{8.47} & \textbf{46.19} & 52.47 & \textbf{3.85} & \textbf{42.37} & \textbf{50.97} & 1.85 \\
SmartDJ$^*$ w/o planer & 42.40 & {55.77} & 1.64 & 33.99 & 44.28 & 3.39 & 33.27 & 39.05 & 1.92 & 34.82 & 35.34 & \textbf{3.70} \\
SmartDJ$^*$ w/ planer & 48.72 & \textbf{63.67} & \textbf{8.20} & 38.30 & 39.00 & 5.08 & 37.05 & 46.09 & 0.00 & 29.61 & 46.57 & 0.00 \\

\bottomrule
\bottomrule
\end{tabular}
}

\end{table*}

\section{Experimental Setup}

\subsection{Benchmarking Candidates}
We evaluate five recent audio editing models on the MMAE benchmark: Step-Audio-EditX~\cite{yan2025Step-Audio-EditX}, Ming-UniAudio~\cite{yan2025Ming-UniAudio}, MMEdit~\cite{tao2025mmedit}, Audio-Omni~\cite{tian2026Audio-Omni}, and SmartDJ~\cite{lan2025SmartDJ}. All models are end-to-end systems, except that SmartDJ is additionally tested with an external planner (Gemini 2.0 Flash~\cite{google_gemini_update_2024}) that decomposes complex instructions into sequential atomic edits. We denote these two settings as SmartDJ w/o planner and SmartDJ w/ planner, respectively.
Due to input length constraints, MMEdit, Audio-Omni, and SmartDJ are evaluated only on samples with input duration $\leq$10 seconds (801 out of 2,000 samples). Step-Audio-EditX and Ming-UniAudio are evaluated on the full set.
In addition, we include two reference baselines to contextualize model performance:
\begin{itemize}
    \item \textit{Identity}: directly returns the input audio without any modification, representing an upper bound on consistency but a lower bound on instruction following.
    \item \textit{Noise}: outputs pure Gaussian noise of matching duration, representing a baseline where no meaningful content is preserved.
\end{itemize}

\subsection{Evaluation Details}
We employ Qwen3-Omni~\cite{xu2025Qwen3-omni} as the external MLLM judger. For each rubric, the judger is provided with the relevant audio(s) and prompted to perform explicit perception and reasoning before selecting an answer from the given options. 
To ensure evaluation stability, each rubric is queried three times independently; a binary score of 1 is awarded via a majority vote (at least 2/3 alignment with the ground truth) and 0 otherwise. Option positions are randomly shuffled per query to mitigate positional bias. 
For each sample, rubric scores are averaged within their respective axes to yield the \textit{Instruction Following Rate} (IFR) and \textit{Consistency Rate} (CR), respectively. We additionally report the \textit{Exact Match Rate} (EMR), defined as the proportion of samples where all rubrics are answered correctly, which serves as a stringent metric quantifying the ratio of perfectly executed edits. 
\section{Experimental Results}

\subsection{Main Results}
\label{subsec:main-results}
The main evaluation results are presented in Table~\ref{tab:main_results}. 
Across all evaluated models, the EMR consistently remains below 5\%, with several models even hitting absolute zero (0\%) in complex mixed-modality settings, confirming that MMAE poses substantial challenges for current audio editing systems.
In the full evaluation set, Step-Audio-EditX establishes the top baseline yet still only achieves a modest 44.86\% IFR and 58.88\% CR, while Ming-UniAudio drops significantly lower to 29.82\% IFR and 52.71\% CR. Among the models restricted to the $\leq$10s subset, Audio-Omni leads with 50.73\% IFR and 56.93\% CR, followed by MMEdit and the SmartDJ variants.

When breaking down performance by task complexity (Table~\ref{tab:result_complexity}), a universal degradation is observed across all models transitioning from single to multiple categories. 
Cross-modal analysis (Table~\ref{tab:result_single_modality} and \ref{tab:result_mix_modality}) further highlights distinct domain biases. Step-Audio-EditX and Ming-UniAudio perform relatively better on speech, while Audio-Omni shows better results on sound and music editing tasks. For mixed-modality tasks, all models exhibit notably lower scores, with Sound-Music-Speech being the hardest category.

Overall, these results indicate that current models possess basic audio editing capabilities, yet consistently fail to achieve flawless edits. They either miss certain intended modifications (low IFR), inadvertently alter content that should be preserved (low CR), and hardly demonstrate the ability to perform perfect editing (low EMR).

\subsection{Observation \& Discussion}

We highlight several key findings below.

\textbf{1) Higher complexity and mixed modalities degrade performance.}
As shown in Table~\ref{tab:result_complexity}, all models exhibit a clear performance drop from single to multiple complexity tasks. For example, Audio-Omni's IFR decreases from 58.43\% to 41.70\%, and its CR drops from 64.57\% to 47.94\%. Similarly, comparing Table~\ref{tab:result_single_modality} and Table~\ref{tab:result_mix_modality}, mixed-modality tasks are generally harder: the sound-music-speech category consistently yields the lowest scores across models, while speech editing in isolation tends to produce the highest CR (e.g., 77.27\% for Step-Audio-EditX). 
These findings indicate that current models lack structural robustness for editing tasks that require complex reasoning and cross-domain synchronization, highlighting the critical need to bridge the gap between reactive single-operation edits and universal, mixed-modality audio manipulation.

\textbf{2) IFR and CR present a fundamental trade-off.}
The reference baselines illustrate this clearly. The Identity baseline achieves near-perfect CR (94.13\%) but poor IFR (27.37\%). Its non-trivial IFR partly comes from extraction tasks or multi-round scenarios where the final output happens to match the original. The Noise baseline attains an IFR of 32.08\% but an extremely low CR of 15.68\%, as random noise can accidentally satisfy certain deletion-verification rubrics. Comparing these baselines to the evaluated models further highlights the inadequacy of current systems: effective audio editing requires simultaneously making precise modifications and preserving unrelated content, a balance that remains difficult. This also verifies our design of reporting IFR and CR separately rather than as a single composite score since a combined metric would allow models to trivially inflate consistency by simply not editing. The EMR metric complements this by measuring the proportion of samples where all dimensions are fully satisfied.

\textbf{3) Average competency and flawless execution show divergence.}
Interestingly, we discover a clear decoupling between average dimension scores (IFR \& CR) and perfect editing rates (EMR). Intuitively, a model with higher average sub-metrics should yield a superior exact match rate. However, our empirical results challenge this assumption: Step-Audio-EditX substantially outperforms Ming-UniAudio in both average IFR (44.86\% vs.\ 29.82\%) and CR (58.88\% vs.\ 52.71\%), yet its EMR is unexpectedly lower (3.05\% vs.\ 3.20\%). 
This discrepancy conceptually mirrors the ``mean-seeking'' versus ``mode-seeking'' behaviors in generative modeling. Step-Audio-EditX acts as a mean-seeking generalist editor: it achieves broad competence by partially satisfying instructions across samples, but it frequently makes minor errors that ruin the perfect edit, leading to a low EMR. 
Conversely, Ming-UniAudio exhibits as a mode-seeking specialist model: it fails entirely on a large portion of the data, but when it does succeed, it hits the exact target, resulting in a higher EMR. 
This highlights that optimizing for average metric improvements does not linearly guarantee holistic reliability for true audio editing intelligence. 

\textbf{4) Agent-guided planning shows limited improvement.}
Comparing SmartDJ with and without its planner, we observe no consistent improvement. The planner variant achieves higher IFR (42.26\% vs.\ 38.20\%) but lower CR (48.33\% vs.\ 55.41\%), and does not outperform SmartDJ w/o planner on overall EMR. Further error analysis reveals that this underperformance stems from bottlenecks in both understanding and generation. On the understanding side, the external planner still struggles with precise multimodal perception, often misinterpreting complex audio contexts. On the generation side, the base model cannot reliably execute atomic operations. Consequently, while decomposing tasks marginally improves instruction adherence (higher IFR), forcing a fragile base model through cascaded, iterative generation steps inevitably accumulates artifacts and severely degrades audio consistency (lower CR). This suggests that future research should focus on building up the base model's editing fidelity before relying purely on symbolic high-level planners. 
\section{Conclusion}
\label{sec:conclusion}

In this work, we present MMAE, a Massive Multitask Audio Editing benchmark. Notably, it is the first comprehensive benchmark for evaluating instruction-guided audio editing across sound, music, speech, and their mix. Motivated by the lack of a unified and rigorous evaluation framework in this rapidly evolving field, MMAE introduces a systematic taxonomy spanning modality, complexity, and operation dimensions, paired with a rubric-based evaluation paradigm that enables fine-grained, objective, and scalable assessment of editing quality along both instruction following and content consistency.

Our evaluation of representative audio editing models reveals that current systems, despite possessing basic editing capabilities, remain far from achieving reliable and precise edits. Overall performance is low across all evaluated models, with exact match rates below 5\%, and significant degradation is observed on complex multi-operation tasks and mixed-modality scenarios. These findings underscore the substantial gap between current capabilities and the demands of real-world audio editing applications.

MMAE highlights key directions in this field including improving atomic editing fidelity, developing models with universal modality support, and advancing robust agent-guided systems for compositional editing. We hope MMAE serves as an effective, challenging yet inspiring benchmark and resource for the community to track progress, identify bottlenecks, and guide future research in audio editing.

\clearpage
\bibliographystyle{unsrtnat}
\bibliography{main}

\clearpage
\begin{appendices}

\section{Demo Examples}

We present representative samples from the MMAE benchmark to illustrate the diversity of tasks and the granularity of rubric-based evaluation.

\vspace{8pt}
\noindent\rule{\textwidth}{0.5pt}
\vspace{4pt}

\noindent\textbf{Case 1} \hfill \textit{multi-audio} $\mid$ \textit{music} $\mid$ \textit{Global -- Foreground change}

\noindent\textbf{Instruction:} Change all the words in the lyrics of audio2 to ``Hachimi`` and use the human voice timbre of audio1.

\vspace{4pt}

\begin{tabularx}{\textwidth}{|c|c|X|}
\hline
\textbf{\#} & \textbf{Category} & \textbf{Rubric} \\
\hline
1 & IF & \textbf{Q:} In <audio output>, what are the words mainly sung clearly by the lead vocalist? \newline \textcolor{teal}{\cmark\ Repeatedly sung onomatopoeic word ``Hachimi``} \newline \xmark\ Mainly German lyrics (such as ``Zuerst lag ich in einem Ei``) \newline \xmark\ Repeatedly sung Japanese ``はちみつ`` \newline \xmark\ Cannot hear the specific words clearly (like humming / no lyrics) \newline \xmark\ Other lyrics, or lyrics mixing multiple languages \newline \xmark\ None of the above \\
\hline
2 & IF & \textbf{Q:} Does the timbre of the human voice in <audio output> sound like <audio input1>? \newline \textcolor{teal}{\cmark\ Yes} \newline \xmark\ No \newline \xmark\ No human voice heard \newline \xmark\ None of the above \\
\hline
3 & Con. & \textbf{Q:} Compare <audio output> and <audio input2>: is the overall accompaniment basically the same? \newline \textcolor{teal}{\cmark\ Yes, the overall accompaniment is basically the same} \newline \xmark\ No, the accompaniment is clearly changed / replaced / has elements greatly added or removed \newline \xmark\ No accompaniment \newline \xmark\ None of the above \\
\hline
4 & Con. & \textbf{Q:} Compare <audio output> and <audio input2>. Are the melody and rhythm of the vocals basically the same? \newline \textcolor{teal}{\cmark\ Yes, basically consistent} \newline \xmark\ No \newline \xmark\ No vocals \newline \xmark\ None of the above \\
\hline
5 & Con. & \textbf{Q:} Compare <audio output> and <audio input2>: Does <audio output> exhibit obvious audio quality degradation (such as newly added background noise/electrical hum, obvious distortion clipping, metallic compression artifacts, or pumping that becomes suddenly louder and quieter)? \newline \textcolor{teal}{\cmark\ No, no obvious audio quality degradation is heard} \newline \xmark\ Yes, obvious audio quality degradation can be heard \newline \xmark\ None of the above \\
\hline
\end{tabularx}

\vspace{8pt}
\noindent\rule{\textwidth}{0.5pt}
\vspace{4pt}

\noindent\textbf{Case 2} \hfill \textit{multi-round} $\mid$ \textit{speech} $\mid$ \textit{Local -- Replacement}

\noindent\textbf{Round 1 Instruction:} Change the order of the first and second authors mentioned in the commit.

\noindent\textbf{Round 2 Instruction:} Change the order of the second and third authors mentioned in the commit.

\vspace{4pt}

\begin{tabularx}{\textwidth}{|c|c|X|}
\hline
\textbf{\#} & \textbf{Category} & \textbf{Rubric} \\
\hline
1 & IF & \textbf{Q:} In <audio output[0s:]>, in this continuous reading aloud of names, which of the following best matches the order of the first three names read? \newline \textcolor{teal}{\cmark\ Hongyi Li → Shinji Watanabe → Abdulrahim Mohammed} \newline \xmark\ Abdulrahim Mohammed → Hongyi Li → Shinji Watanabe \newline \xmark\ Hongyi Li → Abdulrahim Mohammed → Shinji Watanabe \newline \xmark\ Shinji Watanabe → Hongyi Li → Abdulrahim Mohammed \newline \xmark\ None of the above \\
\hline
2 & IF & \textbf{Q:} Compare <audio output[2.15s:4.9s]> and <audio input1[4.35s:7.6s]>: Is the name that is read aloud first in the <audio output[2.15s:4.9s]> segment the same name content as ``Me Hongyi Li,`` in the <audio input1[4.35s:7.6s]> segment? \newline \textcolor{teal}{\cmark\ Yes, both are ``Me Hongyi Li``} \newline \xmark\ No, it is not this name or it is unclear \newline \xmark\ None of the above \\
\hline
3 & IF & \textbf{Q:} Compare <audio output[4.4s:7.9s]> and <audio input1[6.6s:9.8s]>: Is the name read immediately afterward in the <audio output[4.4s:7.9s]> segment the same name content as “Shinji Watanabe,” in the <audio input1[6.6s:9.8s]> segment? \newline \textcolor{teal}{\cmark\ Yes, both are “Shinji Watanabe”} \newline \xmark\ No, it is not that name or it is unclear \newline \xmark\ None of the above \\
\hline
4 & IF & \textbf{Q:} Compare <audio output[6.8s:10.3s]> and <audio input1[2.15s:5.35s]>: Is the name subsequently read aloud in the <audio output[6.8s:10.3s]> segment the same name content as ``Abdulrahim Mohammed,`` in the <audio input1[2.15s:5.35s]> segment? \newline \textcolor{teal}{\cmark\ Yes, both are ``Abdulrahim Mohammed``} \newline \xmark\ No, it is not that name or is unclear \newline \xmark\ None of the above \\
\hline
5 & IF & \textbf{Q:} In <audio output[2.15s:9.8s]>, which of the following groups of names can all be clearly heard (order not required), and each appears at least once? \newline \textcolor{teal}{\cmark\ Abdulrahim Mohammed, Hongyi Li, Shinji Watanabe} \newline \xmark\ Tara Sainath, Karen Livescu, Shang-Wen Li \newline \xmark\ Yvonne Dambhar, Emmanuel Dupoux, Tara Sainath \newline \xmark\ Only one or two of them can be heard, and they cannot be identified as a group \newline \xmark\ None of the above \\
\hline
6 & Con. & \textbf{Q:} Compare <audio output[0s:2.7s]> and <audio input1[0s:2.7s]>: Can the same sentence ``And this is the organization committee,`` be heard in both segments, and are the speaker's timbre and tone consistent? \newline \textcolor{teal}{\cmark\ Yes, the content and speaker characteristics are consistent} \newline \xmark\ No, the content is missing/different or the speaker characteristics are clearly different \newline \xmark\ None of the above \\
\hline
7 & Con. & \textbf{Q:} Compare <audio output[1.7s:3.15s]> and <audio input1[1.7s:3.15s]>: Can a softly spoken filler word ``uh,`` (around approximately 00:02.2) be heard in both segments, accompanied by a similar brief pause? \newline \textcolor{teal}{\cmark\ Yes, both segments have a similar ``uh,`` and pause} \newline \xmark\ No, the ``uh,`` is clearly absent or very different in <audio output[1.7s:3.15s]> or <audio input1[1.7s:3.15s]> \newline \xmark\ None of the above \\
\hline
\end{tabularx}

\begin{tabularx}{\textwidth}{|c|c|X|}
\hline
\textbf{\#} & \textbf{Category} & \textbf{Rubric} \\
\hline
8 & Con. & \textbf{Q:} Compare <audio output[2.15s:3.45s]> and <audio input1[2.15s:3.45s]>: Is the word “include,” present in both segments and similar in pronunciation/rhythm? \newline \textcolor{teal}{\cmark\ Yes, include is consistently present and pronounced similarly} \newline \xmark\ No, include is missing/changed into another word/has obvious differences \newline \xmark\ None of the above \\
\hline
9 & Con. & \textbf{Q:} Compare <audio output[8.8s:11.9s]> and <audio input1[8.8s:11.9s]>: Are the names read in the two segments both “Tara Sainath,” (including the feeling of a comma-like pause at the end)? \newline \textcolor{teal}{\cmark\ Yes, both are “Tara Sainath,”} \newline \xmark\ No, the names are different or unclear \newline \xmark\ None of the above \\
\hline
10 & Con. & \textbf{Q:} Compare <audio output[10.9s:13.7s]> and <audio input1[10.9s:13.7s]>: Are the names read in the two segments both ``Karen Livescu,``? \newline \textcolor{teal}{\cmark\ Yes, both are ``Karen Livescu,``} \newline \xmark\ No, the names are different or unclear \newline \xmark\ None of the above \\
\hline
11 & Con. & \textbf{Q:} Compare <audio output[12.7s:15.1s]> and <audio input1[12.7s:15.1s]>: are the names read aloud in the two segments both ``Shang-Wen Li,``? \newline \textcolor{teal}{\cmark\ Yes, both are ``Shang-Wen Li,``} \newline \xmark\ No, the names are different or unclear \newline \xmark\ None of the above \\
\hline
12 & Con. & \textbf{Q:} Compare <audio output[14.1s:15.8s]> and <audio input1[14.1s:15.8s]>: In both segments, can ``Yvonne Dambhar,`` be heard, immediately followed by a soft ``uh,`` and then entering the final name? \newline \textcolor{teal}{\cmark\ Yes, the structure and content are consistent (Yvonne Dambhar → uh → final name)} \newline \xmark\ No, some part is missing or the order/content is clearly different \newline \xmark\ None of the above \\
\hline
13 & Con. & \textbf{Q:} Compare the entire <audio output> and <audio input1>: Is the overall audio quality of <audio output> not significantly degraded (e.g., no newly introduced distortion, obvious noise floor, pops, clipping, overly strong compression / metallic artifacts)? \newline \textcolor{teal}{\cmark\ Yes, no obvious newly introduced audio-quality degradation is heard} \newline \xmark\ No, obvious newly introduced audio-quality degradation can be heard \newline \xmark\ None of the above \\
\hline
\end{tabularx}

\vspace{8pt}
\noindent\rule{\textwidth}{0.5pt}
\vspace{4pt}

\noindent\textbf{Case 3} \hfill \textit{multi-hop} $\mid$ \textit{sound} $\mid$ \textit{Local -- Removal}

\noindent\textbf{Instruction:} Remove barks from younger dogs.

\vspace{4pt}

\begin{tabularx}{\textwidth}{|c|c|X|}
\hline
\textbf{\#} & \textbf{Category} & \textbf{Rubric} \\
\hline
1 & IF & \textbf{Q:} In <audio output>, can clear dog barking / dog yelping transient sounds (short, sharp ``woof / whine`` type) be heard? \newline \textcolor{teal}{\cmark\ Can be heard} \newline \xmark\ Cannot be heard \newline \xmark\ None of the above \\
\hline
2 & Con. & \textbf{Q:} In <audio output>, can multiple short dog barks (possibly slightly overlapping) be heard? \newline \textcolor{teal}{\cmark\ Multiple dog barks / barking can be heard} \newline \xmark\ No dog barking can be heard; overall it is close to silence \newline \xmark\ None of the above \\
\hline
\end{tabularx}

\begin{tabularx}{\textwidth}{|c|c|X|}
\hline
\textbf{\#} & \textbf{Category} & \textbf{Rubric} \\
\hline
3 & IF & \textbf{Q:} In <audio output>, is it possible to hear one or two short dog barks / dog vocalizations at the end? \newline \textcolor{teal}{\cmark\ Can be heard; there is still dog barking / dog vocalization at the end} \newline \xmark\ Cannot be heard; there is no dog barking / dog vocalization at the end \newline \xmark\ None of the above \\
\hline
4 & Con. & \textbf{Q:} Compare the overall background of <audio output> and <audio input1> (regardless of whether there is dog barking). Do both present near-complete silence, with no added background noise/hum/ambient sound? \newline \textcolor{teal}{\cmark\ Yes, the backgrounds of both are similarly close to silent and have no added sounds} \newline \xmark\ No, there is a clear difference in background noise between <audio output> and <audio input1>, or added ambient sound appears \newline \xmark\ None of the above \\
\hline
5 & Con. & \textbf{Q:} Compare <audio output> and <audio input1>. Does <audio output> contain any new non-canine sounds that are not present in <audio input1> (such as cat meowing, bird calls, human voices, music, or mechanical sounds)? \newline \textcolor{teal}{\cmark\ No, no new non-canine sounds appeared} \newline \xmark\ Yes, at least one new non-canine sound appeared \newline \xmark\ None of the above \\
\hline
6 & Con. & \textbf{Q:} Compare the overall audio quality of <audio output> and <audio input1>. Does <audio output> show noticeable degradation (such as distortion, clipping, over-compression/pumping, obvious quantization noise, or high-frequency graininess)? \newline \textcolor{teal}{\cmark\ No, <audio output> does not show noticeable audio quality degradation} \newline \xmark\ Yes, <audio output> shows noticeable audio quality degradation \newline \xmark\ None of the above \\
\hline
7 & IF & \textbf{Q:} Does <audio output> contain only one type of dog barking sound? \newline \textcolor{teal}{\cmark\ Yes} \newline \xmark\ No, it contains at least 2 types of dog barking sounds. \newline \xmark\ None of the above \\
\hline
8 & IF & \textbf{Q:} Which type of dog bark in <audio input1> is the dog barking sound in <audio output> consistent with? \newline \textcolor{teal}{\cmark\ The sharp, thin, clear bark of a young dog.} \newline \xmark\ The deep, melodious bark of an older dog. \newline \xmark\ None of the above \\
\hline
\end{tabularx}

\vspace{8pt}
\noindent\rule{\textwidth}{0.5pt}
\vspace{4pt}

\noindent\textbf{Case 4} \hfill \textit{multi-part} $\mid$ \textit{sound-music-speech} $\mid$ \textit{Global -- Foreground change}

\noindent\textbf{Instruction:} Change the accented Chinese dialogue in this clip to standard Mandarin pronunciation.

\vspace{4pt}

\begin{tabularx}{\textwidth}{|c|c|X|}
\hline
\textbf{\#} & \textbf{Category} & \textbf{Rubric} \\
\hline
1 & IF & \textbf{Q:} Compare <audio output[0.0s:1.7s]> and <audio input1[0.0s:1.7s]>. In which segment is the male voice's Chinese pronunciation closer to ``Standard Mandarin`` (fewer regional accent features)? \newline \textcolor{teal}{\cmark\ <audio output[0.0s:1.7s]> is closer to Standard Mandarin} \newline \xmark\ <audio input1[0.0s:1.7s]> is closer to Standard Mandarin \newline \xmark\ The two are comparable in degree of Mandarin, with no obvious difference audible \newline \xmark\ None of the above \\
\hline
\end{tabularx}

\begin{tabularx}{\textwidth}{|c|c|X|}
\hline
\textbf{\#} & \textbf{Category} & \textbf{Rubric} \\
\hline
2 & IF & \textbf{Q:} Compare <audio output[11.8s:16.0s]> and <audio input1[11.8s:16.0s]>. In which segment is the male voice's Chinese pronunciation closer to ``Standard Mandarin`` (with fewer regional accent characteristics)? \newline \textcolor{teal}{\cmark\ <audio output[11.8s:16.0s]> is closer to Standard Mandarin} \newline \xmark\ <audio input1[11.8s:16.0s]> is closer to Standard Mandarin \newline \xmark\ No obvious difference can be heard between the two \newline \xmark\ None of the above \\
\hline
3 & Con. & \textbf{Q:} In <audio output[0.0s:1.4s]>, which of the following does this line sound more like? \newline \textcolor{teal}{\cmark\ “我要验牌”} \newline \xmark\ Not this sentence (the words have clearly changed or the content is inconsistent) \newline \xmark\ Unclear / covered by other sounds, making it impossible to judge \newline \xmark\ None of the above \\
\hline
4 & Con. & \textbf{Q:} In <audio output[11.8s:14.5s]>, which of the following does this line sound more like? \newline \textcolor{teal}{\cmark\ “牌没有问题”} \newline \xmark\ Not this line (the words have clearly changed or the content is inconsistent) \newline \xmark\ Unclear / covered by other sounds, making it impossible to judge \newline \xmark\ None of the above \\
\hline
5 & IF & \textbf{Q:} Compare <audio output[0.0s:1.4s]> and <audio input1[0.0s:1.4s]>. Do the two sound like the same adult male voice speaking (timbre and identity characteristics basically consistent), rather than having changed to another speaker? \newline \textcolor{teal}{\cmark\ They are the same male voice (timbre/identity basically consistent)} \newline \xmark\ No, it sounds like a different person (gender/age/timbre clearly different) \newline \xmark\ Unable to determine \newline \xmark\ None of the above \\
\hline
6 & IF & \textbf{Q:} Compare <audio output[11.8s:14.5s]> and <audio input1[11.8s:14.5s]>. Do they sound like the same adult male voice speaking (with basically consistent timbre and identity characteristics), rather than having switched to another speaker? \newline \textcolor{teal}{\cmark\ The same male voice (timbre/identity basically consistent)} \newline \xmark\ No, it sounds like it switched to another person (gender/age/timbre clearly different) \newline \xmark\ Unable to determine \newline \xmark\ None of the above \\
\hline
7 & Con. & \textbf{Q:} Compare <audio output[0.0s:1.4s]> and <audio input1[0.0s:1.4s]>. Are their speaking rate and pause positions basically consistent (disregarding accent differences, only considering whether rhythm/duration/start and end points are aligned)? \newline \textcolor{teal}{\cmark\ Basically consistent (rhythm and start/end points are about the same)} \newline \xmark\ Inconsistent (clearly faster/slower, lengthened/shortened, or alignment points changed) \newline \xmark\ Cannot determine \newline \xmark\ None of the above \\
\hline
8 & Con. & \textbf{Q:} Compare <audio output[11.8s:14.5s]> and <audio input1[11.8s:14.5s]>. Are their speaking rate and pause positions basically consistent (disregarding accent differences, only considering whether rhythm/duration/start and end points are aligned)? \newline \textcolor{teal}{\cmark\ Basically consistent (rhythm and start/end points are about the same)} \newline \xmark\ Inconsistent (clearly faster/slower, lengthened/shortened, or alignment points changed) \newline \xmark\ Unable to determine \newline \xmark\ None of the above \\
\hline
\end{tabularx}

\begin{tabularx}{\textwidth}{|c|c|X|}
\hline
\textbf{\#} & \textbf{Category} & \textbf{Rubric} \\
\hline
9 & Con. & \textbf{Q:} Compare <audio output[0.4s:12.5s]> and <audio input1[0.4s:12.5s]>. Are the close-up plastic/cellophane rubbing sounds (timing of occurrence, changes in density, and overall intensity trend) basically the same? \newline \textcolor{teal}{\cmark\ Basically the same (sounds like the same segment of rubbing sound)} \newline \xmark\ Not the same (replaced / clearly distorted / different intensity trend) \newline \xmark\ The rubbing sound in <audio output[0.4s:12.5s]> is clearly missing, or a clearly different new ambient sound appears \newline \xmark\ None of the above \\
\hline
10 & Con. & \textbf{Q:} Compare <audio output[14.0s:16.0s]> and <audio input1[14.0s:16.0s]>. Is the quiet tail segment at the end (residual reverberation and slight background noise) basically consistent, with no additional newly added sound effects or human voices? \newline \textcolor{teal}{\cmark\ Basically consistent and no newly added elements} \newline \xmark\ Inconsistent: newly added sound effects/human voices appear, or the background noise/tail sound is noticeably altered \newline \xmark\ Unable to determine \newline \xmark\ None of the above \\
\hline
11 & Con. & \textbf{Q:} Compare the entire <audio output> with the entire <audio input1>. Does the overall audio quality of <audio output> show degradation (such as obvious distortion/clipping, watery sounds caused by excessive noise reduction, obvious compression pumping, an abnormal increase in noise, or a significant decrease in clarity)? \newline \textcolor{teal}{\cmark\ No obvious degradation heard} \newline \xmark\ Obvious degradation heard \newline \xmark\ Unable to determine \newline \xmark\ None of the above \\
\hline
\end{tabularx}

\vspace{8pt}
\noindent\rule{\textwidth}{0.5pt}
\vspace{4pt}

\noindent\textbf{Case 5} \hfill \textit{multi-instruction} $\mid$ \textit{music-speech} $\mid$ \textit{Global -- Background change + Global -- Alteration}

\noindent\textbf{Instruction:} Change the background music to a guitar with the exactly same melody, while making the vocals deeper and more resonant, without changing the spoken content.

\vspace{4pt}

\begin{tabularx}{\textwidth}{|c|c|X|}
\hline
\textbf{\#} & \textbf{Category} & \textbf{Rubric} \\
\hline
1 & IF & \textbf{Q:} Compare the main accompaniment melody carrier in <audio input1[0.0s:10.0s]> and <audio output[0.0s:10.0s]>. Which one more clearly exhibits the characteristics of ``guitar plucking / strumming / acoustic guitar or electric guitar timbre``? \newline \textcolor{teal}{\cmark\ <audio output[0.0s:10.0s]> is more clearly dominated by guitar timbre} \newline \xmark\ <audio input1[0.0s:10.0s]> is more clearly dominated by guitar timbre \newline \xmark\ Neither is clearly dominated by guitar timbre \newline \xmark\ None of the above \\
\hline
2 & IF & \textbf{Q:} At the melodic level, compare <audio input1[0.0s:10.0s]> and <audio output[0.0s:10.0s]>: do the ``pitch direction and rhythmic contour`` of the main melody/motif of the two sound basically consistent (like the same melody presented by different instruments)? \newline \textcolor{teal}{\cmark\ Basically consistent} \newline \xmark\ Clearly inconsistent (the melody/motif has clearly changed) \newline \xmark\ The melody in this segment of <audio output[0.0s:10.0s]> or <audio input1[0.0s:10.0s]> is unclear, making it difficult to compare \newline \xmark\ None of the above \\
\hline
\end{tabularx}

\begin{tabularx}{\textwidth}{|c|c|X|}
\hline
\textbf{\#} & \textbf{Category} & \textbf{Rubric} \\
\hline
3 & IF & \textbf{Q:} Compare the perceived pitch of the male narration voice in <audio input1[0.0s:10.0s]> and <audio output[0.0s:10.0s]>. Which one is more ``deep`` (overall pitch is lower)? \newline \textcolor{teal}{\cmark\ <audio output[0.0s:10.0s]> is deeper} \newline \xmark\ <audio input1[0.0s:10.0s]> is deeper \newline \xmark\ The difference between the two is not obvious \newline \xmark\ None of the above \\
\hline
4 & IF & \textbf{Q:} In <audio output[0.7s:4.0s]>, can the narration's spoken content still be recognized as ``最高の伪装とは、自らを欺くこと。``? \newline \textcolor{teal}{\cmark\ Yes, it can be recognized as this sentence} \newline \xmark\ No, it can be recognized as different content/sentence \newline \xmark\ The human voice is too unclear to recognize \newline \xmark\ None of the above \\
\hline
5 & IF & \textbf{Q:} In <audio output[5.5s:9.3s]>, can the narrator's spoken content still be recognized as ``俺は教練部の書記官、アルハイゼンだ。``? \newline \textcolor{teal}{\cmark\ Yes, it can be recognized as this sentence} \newline \xmark\ No, it can be recognized as different content / a different sentence \newline \xmark\ The voice is too unclear to recognize \newline \xmark\ None of the above \\
\hline
6 & Con. & \textbf{Q:} Compare the ending treatment of <audio input1> and <audio output>: do both present a structure where ``near the end, all elements suddenly stop simultaneously, and the ending is very clean`` (rather than an obvious fade-out or a very long trailing tail sound)? \newline \textcolor{teal}{\cmark\ Yes, the ending structures of both are consistent} \newline \xmark\ No, the ending structures of <audio output> and <audio input1> are clearly different \newline \xmark\ None of the above \\
\hline
7 & Con. & \textbf{Q:} Compare <audio input1[3.5s:5.6s]> and <audio output[3.5s:5.6s]>: does the transition from the buildup into a denser, stronger beat occur at a similar time position (without being obviously much earlier/later)? \newline \textcolor{teal}{\cmark\ Yes, the time position of the transition point is roughly consistent} \newline \xmark\ No, the transition point in <audio output[3.5s:5.6s]> is clearly much earlier/later or missing \newline \xmark\ None of the above \\
\hline
8 & Con. & \textbf{Q:} Compare <audio input1[0.0s:1.7s]> and <audio output[0.0s:1.7s]>: Do both beginnings present a relatively sparse build-up and gradually introduce the beat/layers (rather than being fully arranged from the start or having almost no content for a long time)? \newline \textcolor{teal}{\cmark\ Yes, the overall build-up approach at the beginning is consistent} \newline \xmark\ No, the build-up approach at the beginning of <audio output[0.0s:1.7s]> has clearly changed \newline \xmark\ None of the above \\
\hline
9 & Con. & \textbf{Q:} Compare <audio input1[0.7s:3.5s]> and <audio output[0.7s:3.5s]>: Are the timing of the narration's appearance and its duration roughly consistent (not obviously advanced/delayed, truncated, or repeated)? \newline \textcolor{teal}{\cmark\ Yes, the timing and duration are roughly consistent} \newline \xmark\ No, the narration timing in <audio output[0.7s:3.5s]> is obviously inconsistent (such as misaligned/truncated/repeated) \newline \xmark\ None of the above \\
\hline
\end{tabularx}

\begin{tabularx}{\textwidth}{|c|c|X|}
\hline
\textbf{\#} & \textbf{Category} & \textbf{Rubric} \\
\hline
10 & Con. & \textbf{Q:} Compare <audio input1[4.5s:10.0s]> and <audio output[4.5s:10.0s]>: Is the speed of the four-beat beat-hitting (overall perceived speed) kept consistent (without obvious speeding up or slowing down)? \newline \textcolor{teal}{\cmark\ Yes, the perceived speed is consistent} \newline \xmark\ No, the perceived speed of <audio output[4.5s:10.0s]> is clearly different \newline \xmark\ None of the above \\
\hline
11 & Con. & \textbf{Q:} Compare <audio input1> and <audio output>: apart from differences in instrument timbre and vocal timbre, does <audio output> contain significant newly added content not present in <audio input1> (such as added vocal layers, abrupt environmental noise, or additional sound-effect passages)? \newline \textcolor{teal}{\cmark\ No obvious newly added content heard} \newline \xmark\ Obvious newly added content heard \newline \xmark\ None of the above \\
\hline
12 & Con. & \textbf{Q:} Compare the overall listening impression of <audio input1> and <audio output>: does <audio output> exhibit obvious audio quality degradation (such as heavier background noise / hum, distortion or clipping, obvious compression pumping, harsh pops, or high-frequency graininess), causing a decrease in clarity? \newline \textcolor{teal}{\cmark\ No obvious audio quality degradation} \newline \xmark\ Obvious audio quality degradation \newline \xmark\ None of the above \\
\hline
\end{tabularx}

\vspace{8pt}
\noindent\rule{\textwidth}{0.5pt}
\vspace{4pt}

\noindent\textbf{Case 6} \hfill \textit{single} $\mid$ \textit{sound-speech} $\mid$ \textit{Local -- Extraction}

\noindent\textbf{Instruction:} Isolate and extract all sounds produced by dogs, such as barking or whining, while suppressing human speech and other environmental background noises.

\vspace{4pt}

\begin{tabularx}{\textwidth}{|c|c|X|}
\hline
\textbf{\#} & \textbf{Category} & \textbf{Rubric} \\
\hline
1 & IF & \textbf{Q:} In <audio output>, can any human speech content be clearly heard (recognizable words or sentences)? \newline \textcolor{teal}{\cmark\ No, basically no recognizable words or sentences can be heard} \newline \xmark\ Yes, human speech words or sentences can be recognized \newline \xmark\ None of the above \\
\hline
2 & IF & \textbf{Q:} In <audio output[4.0s:5.7s]>, is it possible to hear an obvious human shouting/speaking timbre (it is not required to make out the specific words)? \newline \textcolor{teal}{\cmark\ Cannot hear obvious human shouting/speaking} \newline \xmark\ Can hear obvious human shouting/speaking \newline \xmark\ None of the above \\
\hline
3 & IF & \textbf{Q:} In <audio output[2.0s:15.0s]>, can obvious canine vocalizations such as dog barking / dog howling / whimpering still be heard? \newline \textcolor{teal}{\cmark\ Yes, obvious canine vocalizations can be heard} \newline \xmark\ No, no obvious canine vocalizations can be heard \newline \xmark\ None of the above \\
\hline
4 & IF & \textbf{Q:} In <audio output[5.0s:10.0s]>, which of the following categories does the dominant sound best fit? \newline \textcolor{teal}{\cmark\ Canine long howling / barking} \newline \xmark\ Human speech / shouting \newline \xmark\ Mainly environmental noise or nearly silent \newline \xmark\ None of the above \\
\hline
\end{tabularx}

\clearpage
\section{Final Meta-Data Format}
\label{sec:metadata-format}

Each sample in the MMAE benchmark is stored as a JSON object. Below is a complete example illustrating the final released format:

\begin{lstlisting}[
  language=json,
  caption={Sample JSON annotation of MMAE.},
  label={lst:meta_data},
  columns=fullflexible,
  keepspaces=true,
  xleftmargin=4pt,
  xrightmargin=4pt,
]
{
    "id": "69e898163a050f39ac567501",
    "complexity": "single",
    "modality": "sound-speech",
    "granularity": [
        "local"
    ],
    "operations": [
        {
            "granularity": "local",
            "operation": "extraction"
        }
    ],
    "messages": [
        {
            "role": "user",
            "content": [
                {
                    "type": "text",
                    "text": "Isolate and extract all sounds produced by dogs, such as barking or whining, while suppressing human speech and other environmental background noises."
                },
                {
                    "type": "audio",
                    "audio_url": "wav/69e898163a050f39ac567501/audio1.wav"
                }
            ]
        }
    ],
    "tags": [
        [
            "Acoustic Event Detection / Sound Event Extraction",
            "Canine Vocalizations (Dog Barking / Whining)",
            "Non-Human Audio Signal",
            "Background Noise & Speech Suppression"
        ]
    ],
    "rubrics": [
        {
            "category": "Instruction Following",
            "question": "In <audio output>, can any human speech content be clearly heard (recognizable words or sentences)?",
            "right_choice": "No, basically no recognizable words or sentences can be heard",
            "wrong_choices": [
                "Yes, human speech words or sentences can be recognized",
                "None of the above"
            ]
        },
        {
            "category": "Instruction Following",
            "question": "In <audio output[4.0s:5.7s]>, is it possible to hear an obvious human shouting/speaking timbre (it is not required to make out the specific words)?",
            "right_choice": "Cannot hear obvious human shouting/speaking",
            "wrong_choices": [
                "Can hear obvious human shouting/speaking",
                "None of the above"
            ]
        },
        {
            "category": "Instruction Following",
            "question": "In <audio output[2.0s:15.0s]>, can obvious canine vocalizations such as dog barking / dog howling / whimpering still be heard?",
            "right_choice": "Yes, obvious canine vocalizations can be heard",
            "wrong_choices": [
                "No, no obvious canine vocalizations can be heard",
                "None of the above"
            ]
        },
        {
            "category": "Instruction Following",
            "question": "In <audio output[5.0s:10.0s]>, which of the following categories does the dominant sound best fit?",
            "right_choice": "Canine long howling / barking",
            "wrong_choices": [
                "Human speech / shouting",
                "Mainly environmental noise or nearly silent",
                "None of the above"
            ]
        }
    ]
}
\end{lstlisting}

\clearpage
\section{Evaluation Prompt}
\label{sec:eval_prompt}

Below is the exact prompt used to query the external MLLM judger (Qwen3-Omni) for rubric-based evaluation. The system prompt establishes the role and guidelines, while the user prompt provides task-specific instructions for answering each rubric question.

\begin{AIbox}{System Prompt}
{\small\ttfamily
You are an audio analysis assistant.\par
\vspace{4pt}
Your task is to answer user questions strictly based on the factual content of the provided audio clips. Carefully listen to and analyze the audio before answering.\par
\vspace{4pt}
Audio reference notation:\par
- A bare reference like <audio output> or <audio input1> refers to the full duration of that audio clip.\par
- A reference with [start:end] denotes a time slice of that clip (e.g. <audio output[0.0s:3s]> means the segment from 0.0s to 3.0s).\par
- An omitted start means the beginning of the clip (e.g. <audio input1[:2.0s]> means from the beginning up to 2.0s).\par
- An omitted end means the end of the clip (e.g. <audio output[1.5s:]> means from 1.5s to the end).\par
- A negative value counts backward from the end (e.g. <audio input2[-2.5s:]> means the last 2.5 seconds of that clip).\par
- The slicing semantics follow Python-style [start:end] conventions.\par
\vspace{4pt}
Guidelines:\par
- Base your answers only on information that is clearly present or can be directly inferred from the audio.\par
- Do NOT make up details that are not supported by the audio.\par
- If the audio does not contain enough information to answer the question, say so explicitly.\par
- When relevant, reference specific parts of the audio (e.g., events, sounds, speech content, timing).\par
- Be concise, clear, and accurate.\par
- If the audio contains speech, you may transcribe or summarize relevant portions to support your answer.\par
- If the audio is noisy, ambiguous, or unclear, acknowledge the uncertainty.\par
\vspace{4pt}
Your goal is to provide reliable, evidence-based answers grounded in the audio content only.
}
\end{AIbox}

\vspace{2mm}
\begin{AIbox}{User Prompt}
{\small\ttfamily
<audio \{label1\}>: \{audio1\}\par
\vspace{4pt}

[<audio \{label2\}>: \{audio2\}]\par
\vspace{4pt}

[...]\par
\vspace{4pt}

Based on the objective content of the uploaded audio, answer the following multiple-choice question. First carefully perceive, analyze, and reason about the audio content, then choose exactly one option from the list. Return only JSON with keys ``reason'' and ``choice''. Put the reason first and the final choice last. The choice value must be a single uppercase letter identifying the option.\par
\vspace{4pt}

Question:\par
\{question\}\par
\vspace{4pt}

Choices:\par
\{choices\}
}
\end{AIbox}

\clearpage
\section{Data Curation Platform}
\label{sec:platform}

\begin{figure*}[htbp]
  \centering
  \includegraphics[width=\textwidth]{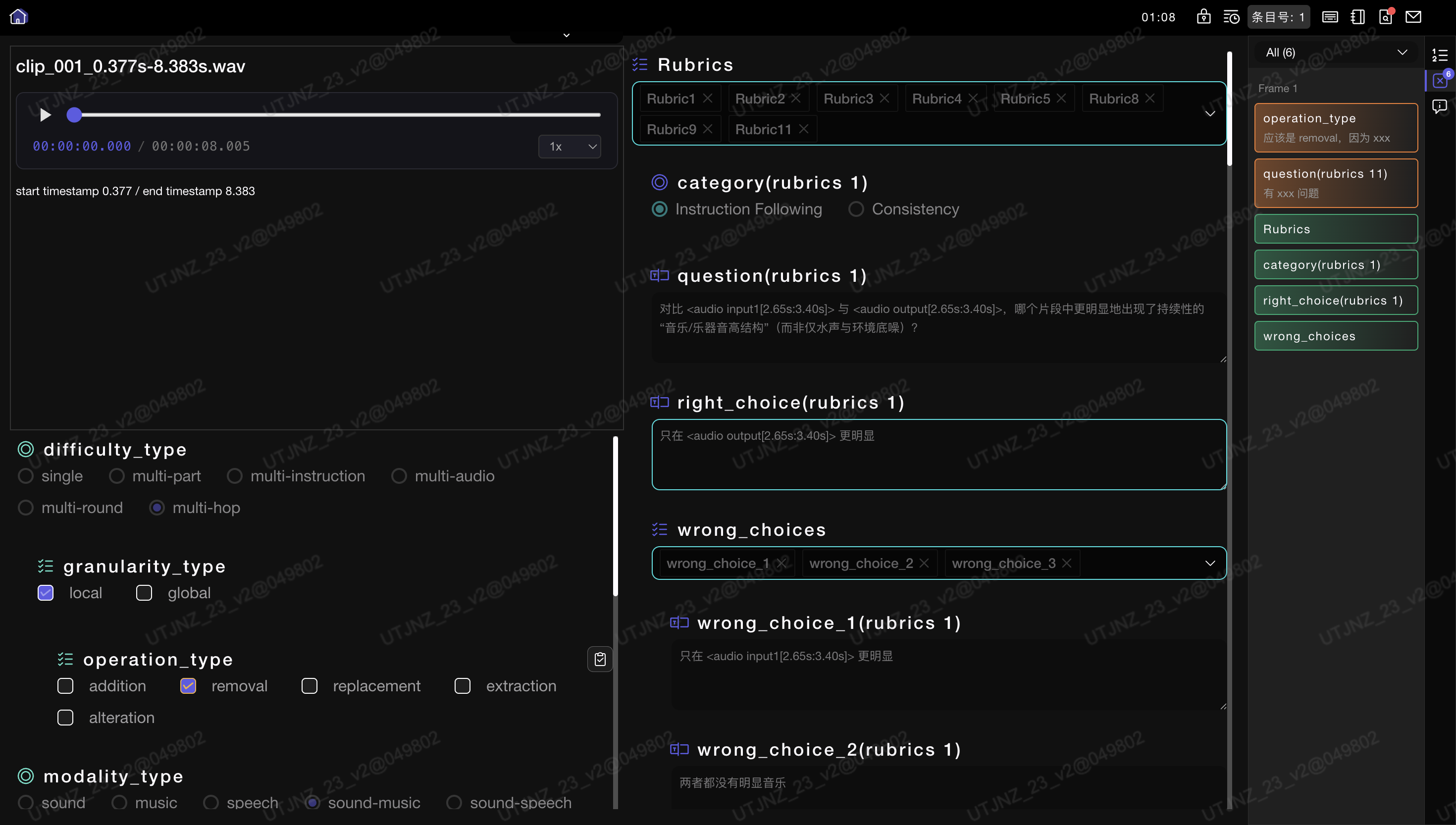}
  \caption{A Snapshot of the platform used for data annotation and quality inspection.}
  \label{fig:platform}
\end{figure*}

We conducted data annotation and quality inspection using a professional platform that supports structured editing, version control, and multi-stage review. All meta data and rubrics were annotated, reviewed and corrected through this system. A Snapshot of the platform interface is shown in Figure~\ref{fig:platform}. 

\end{appendices}

\end{document}